# Anisotropic Thermal Transport in Phase-Transition Layered 2D Alloys WSe$_{2(1-x)}$Te$_{2x}$

Xin Qian[1#], Puqing Jiang[1#], Peng Yu[2#], Xiaokun Gu[3], Zheng Liu[2*] and Ronggui Yang[1*]

[1]Department of Mechanical Engineering, University of Colorado, Boulder, CO 80309

[2]Center for Programmable Materials, School of Materials Science & Engineering, Nanyang Technological University, 639798, Singapore

[3]Institute of Engineering Thermophysics, Shanghai Jiao Tong University, Shanghai, 200240

[*]Email: Ronggui.Yang@Colorado.Edu

Z.Liu@NTU.edu.sg

[#] These authors contribute equally to this work.

**Abstract**

Transition metal dichalcogenide (TMD) alloys have attracted great interests in recent years due to their tunable electronic properties, especially the semiconductor-metal phase transition, along with their potential applications in solid-state memories and thermoelectrics. However, the thermal conductivity of layered two-dimensional (2D) TMD alloys remains largely unexplored despite that it plays a critical role in the reliability and functionality of TMD-enabled devices. In this work, we study the temperature-dependent anisotropic thermal conductivity of the phase-transition 2D TMD alloys WSe$_{2(1-x)}$Te$_{2x}$ in both the in-plane direction (parallel to the basal planes) and the cross-plane direction (along the *c*-axis) using time-domain thermoreflectance measurements. In the WSe$_{2(1-x)}$Te$_{2x}$ alloys, the cross-plane thermal conductivity is observed to be dependent on the heating frequency (modulation frequency of the pump laser) due to the non-equilibrium transport between different phonon modes. Using a two-channel heat conduction model, we extracted the anisotropic thermal conductivity at the equilibrium limit. A clear discontinuity in both the cross-plane and the in-plane thermal conductivity is observed as *x* increases from 0.4 to 0.6 due to the phase transition from the 2H to Td phase in the layered 2D alloys. The temperature dependence of thermal conductivity for the TMD alloys was found to become weaker compared with the pristine 2H WSe$_2$ and Td WTe$_2$ due to the atomic disorder. This work serves as an important starting point for exploring phonon transport in layered 2D alloys.

Transition metal dichalcogenides (TMDs) are a family of layered two-dimensional (2D) materials showing promising applications in electronics,[1-3] photonics,[4-6] hydrogen evolution catalysts,[7-9] and energy conversion and storage.[10-12] Alloys of layered 2D TMDs are of particular interest due to their tunable properties.[13-14] For example, electronic bandgaps can be continuously changed in several TMD alloy systems including $MoS_{2(1-x)}Se_{2x}$,[15-16] $W_xMo_{1-x}S_2$[17-18] and $W_xMo_{1-x}Se_2$.[19] With intrinsically large electronic power factor of TMDs,[20-22] alloying the layered TMDs could potentially lead to highly efficient thermoelectrics where the thermal conductivity can be effectively reduced due to the mass-disorder scattering.[23-24] Along this line, Gu *et al*.[25] performed a first-principles study on the thermal conductivity of $MoS_{2(1-x)}Se_{2x}$ monolayers, and predicted an order-of-magnitude reduction in the thermal conductivity of these alloys as compared with those of the pristine $MoS_2$ and $MoSe_2$. However, there yet exists any experimental work on anisotropic thermal transport in layered TMD alloys, especially as a function of alloy compositions.

Intrigued by the unique electronic structures[26-28] and the abundant phase transition[29-30] in $MoTe_2$ and $WTe_2$, $MoTe_2$ or $WTe_2$ based TMD alloys have also attracted intensive research, especially on the manipulation of the physical properties through controlling the phase transitions. For example, the topological electronic states of $W_xMo_{1-x}Te_2$ can be effectively manipulated by the alloys composition.[31] The hysteresis of the cross-plane thermal conductivity during the 1T'-to-Td phase transition of $W_xMo_{1-x}Te_2$ is also observed, and the phase change point can be tuned to room temperature,[32] which makes it a promising material for phase change memory devices.[33] Depending on the composition, $WSe_{2(1-x)}Te_{2x}$ also exhibits a 2H-to-Td phase transition, which can be used for bandgap tuning.[34] Such phase transition in ternary TMD alloys could be potentially exploited for optimizing thermoelectric performances.[35] However, there exists no study on how such a phase transition would affect the thermal transport properties.

In this Letter, we study the composition-dependent and temperature dependent anisotropic thermal conductivity of the ternary alloy of layered 2D $WSe_{2(1-x)}Te_{2x}$ using an integrated experimental-theoretical approach with the time-domain thermoreflectance (TDTR) measurements. Figure 1a-b shows that atomic structure of the ternary TMD alloy $WSe_{2(1-x)}Te_{2x}$ can be in both 2H phase and Td phase. When the Se atom occupies the majority of the chalcogen sites, the TMD alloy shows a hexagonal 2H phase. A phase transition to the orthorhombic Td phase would happen when Te atom becomes the majority. We prepared single crystalline samples of the layered 2D TMD alloys for the measurement of composition dependent thermal conductivity. The composition fraction $x$ of the $WSe_{2(1-x)}Te_{2x}$ crystals is varied from zero to one, including the pure $WSe_2$ ($x=0$), $WSe_{1.8}Te_{0.2}$ ($x=0.1$), $WSe_{1.6}Te_{0.4}$ ($x=0.2$), $WSe_{1.2}Te_{0.8}$ ($x=0.4$), $WSe_{0.8}Te_{1.2}$ ($x=0.6$), $WSe_{0.4}Te_{1.6}$ ($x=0.8$), and $WTe_2$ ($x=1.0$). These samples are synthesized by chemical vapor transport in a one-step process and the details of synthesis are described in Supporting Information S1. Typical structure of the 2H and Td alloy samples are shown in Figure 1c ($WSe_{1.2}Te_{0.8}$) and Figure 1d ($WSe_{0.4}Te_{1.6}$), respectively. In addition, x-ray diffraction (XRD) and energy dispersive x-ray (EDX) results show that the single crystals of $WSe_{1.2}Te_{0.8}$ (Figure 1e) and $WSe_{0.4}Te_{1.6}$ (Figure 1f) used in this work crystallize in 2H and 1Td structures, respectively, which are consistent with the nominal compositions (Inset, Figure 1e-f).[34]

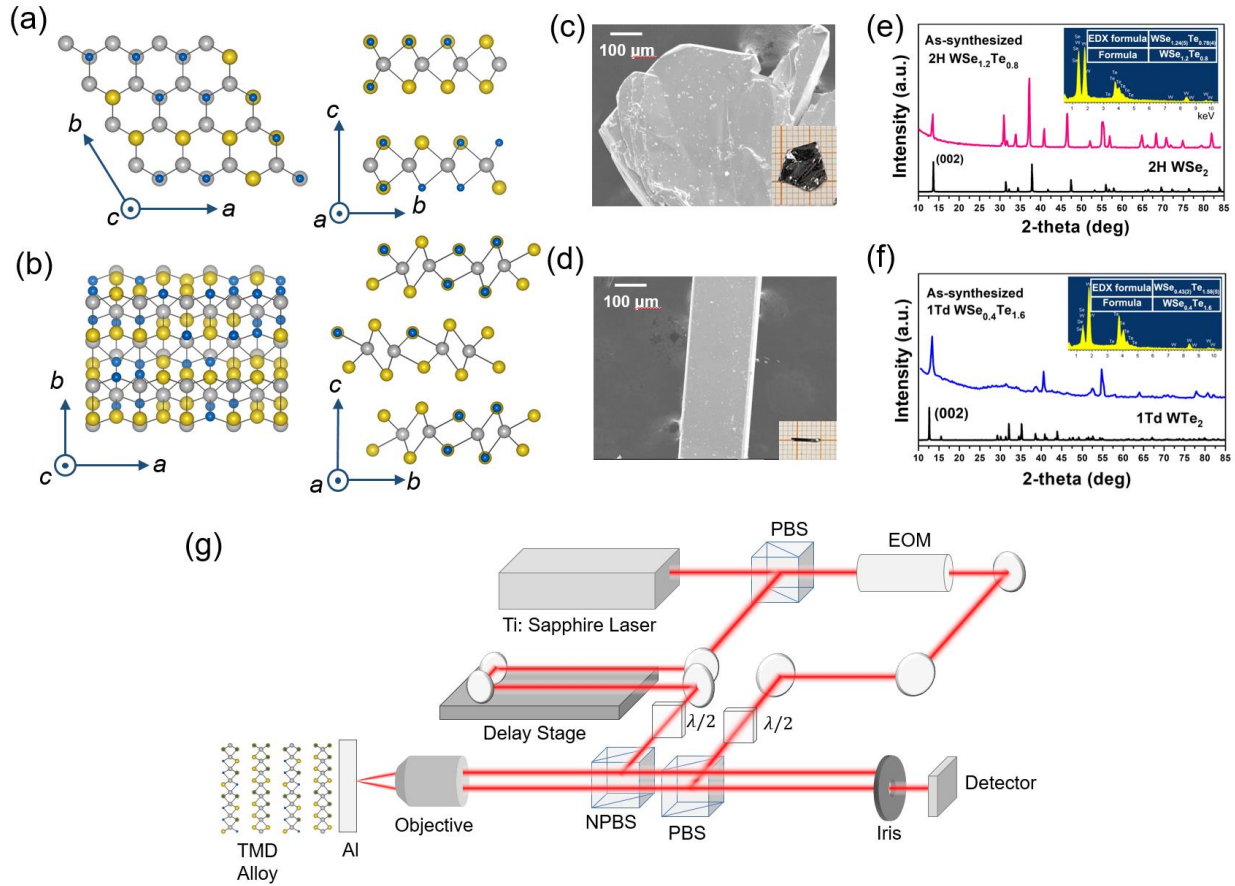

Figure 1. (a) Top view and side view of the structure of the layered 2D TMD alloy in 2H phase. The gray, blue and yellow spheres (not to scale) represent W, Se and Te atoms, respectively. (b) Top view and side view of the structure of the layered 2D TMD alloy in Td phase. (c-d) SEM and optical (inset) images of (c) 2H $WSe_{1.2}Te_{0.8}$ and (d) Td $WSe_{0.4}Te_{1.6}$. The scale bar of SEM images are 100 μm, and each grid in the optical image is 1mm-by-1mm. (e-f) XRD and EDX (inset) characterization of (e) 2H $WSe_{1.2}Te_{0.8}$ and (f) Td phase $WSe_{0.4}Te_{1.6}$. (g) Schematic of the TDTR system used for measuring the anisotropic thermal conductivity. Abbreviations are listed as follows: EOM, electric-optical modulator; PBS/NPBS, polarized/non-polarized beam splitter; $\lambda/2$, half wave plate.

We measure the anisotropic thermal conductivity of layered 2D TMD alloys using a femtosecond laser-based TDTR system as shown in Figure 1g. This system splits the femtosecond laser into a pump beam and a probe beam. The pump beam is modulated by an electric-optical modulator (EOM) at a typical frequency from 0.3 MHz to 10 MHz to create a periodic thermal excitation on the surface of TMD alloy samples, where the surface of TMD alloy samples are deposited with an aluminum metal thin film (~ 75 nm). This Al transducer film absorbs the pump beam and its reflectance changes linearly with the surface temperature rise. The surface temperature change is monitored by detecting the reflected signal of the probe beam which is delayed to arrive at the sample surface by a translational delay stage. The transient surface temperature response is described by the ratio $-V_{in}/V_{out}$ between the in-phase signal $V_{in}$ and the out-of-phase signal $V_{out}$, as shown in Figure 2a. The transient curve is fitted by solving a model for heat conduction in multi-layers using the nonlinear least squared regression to extract the thermal properties.[36-38] Details of our TDTR system along with the data reduction method can be seen in our previous publications.[39-40]

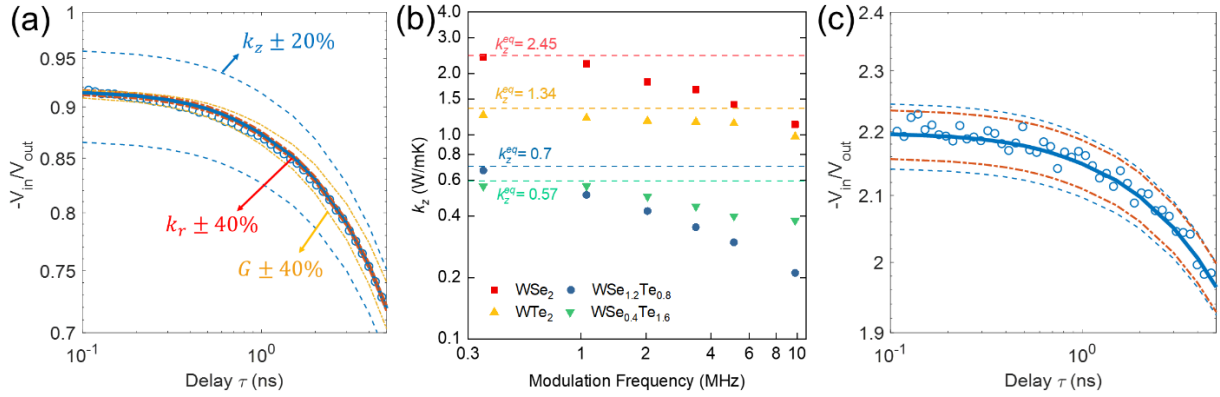

Figure 2. (a) Measured TDTR signal for WSe$_{1.2}$Te$_{0.8}$ using 76 nm Al transducer, a root-mean-square spot radius $w = 8.4$ μm and a modulation frequency $f_0 = 2.16$ MHz. The TDTR signal is dominantly sensitive to $k_z$. The best-fit cross-plane thermal conductivity is $k_z = 0.42$ W/mK. (b) Frequency dependent $k_z$ of WSe$_2$, WSe$_{1.2}$Te$_{0.8}$, WTe$_{1.6}$Se$_{0.4}$ and WTe$_2$. (c) Measuring in-plane thermal conductivity of WTe$_{1.2}$Se$_{0.8}$ sample using $w = 4.2$ μm. The best-fit $k_r = 10.1$ W/mK.

Since the TMD alloys are layered 2D crystals with drastically different in-plane bonding and interlayer bonding, the thermal conductivity should be strongly anisotropic along the in-plane and cross-plane directions. In TDTR, we can separately determine the in-plane and cross-plane thermal conductivity by choosing appropriate operational parameters, i.e. the modulation frequency $f_0$ and the root-mean-squared laser spot radius $w$ between the pump spot radius $w_0$ and probe spot radius $w_1$, i.e., $w = \sqrt{w_0^2 + w_1^2}$.[40] When the laser spot radius $w$ is much larger than the penetration depth $d_p = \sqrt{k_r/\pi C f_0}$ with $k_r$ the in-plane thermal conductivity, $C$ the volumetric heat capacity and $f_0$ the modulation frequency, the temperature field is quasi-one-dimensional in the cross-plane direction, and the TDTR is only sensitive to the cross-plane thermal conductivity $k_z$.[40] In Figure 2a, we show an example of measuring $k_z$ of WSe$_{1.2}$Te$_{0.8}$ with WTe$_2$ fraction $x$ =0.4 using a spot radius $w$ = 8.4 µm. The signal $-V_{in}/V_{out}$ is sensitive to $k_z$ only, especially when the probe delay time is shorter than 1 ns. This allows us to measure $k_z$ independently of $k_r$ and the interface conductance $G$ with the Al transducer. Interestingly, we observed that the measured thermal conductivity $k_z$ decreases as the modulation frequency increases (Figure 2b), if the conventional heat conduction model with one effective thermal conductivity value[36, 38] for the data analysis. Such frequency-dependent $k_z$ was also observed in other layered 2D crystals like pure TMDs[39] and in black phosphorus.[41] Jiang and Qian *et al.* explained this as the non-equilibrium phonon transport between the low frequency and high frequency phonon modes.[39] In the cross-plane direction, the thermal conductivity is mainly contributed by the low frequency phonons with small heat capacity, while the high frequency phonons with large heat capacity have negligible thermal conductivity due to their "flat" phonon bandstructure.[39] Such a large mismatch between heat conductivity and heat capacity of different frequency bands of phonons tends to be in non-equilibrium. When the sample is subjected to periodic laser heating, the heat is accumulated in the high frequency phonons, resulting in their higher temperature than the low frequency phonons. To describe such non-equilibrium phonon transport, we use a two-channel heat conduction model[42] that separates the phonon heat carriers into the high frequency channel and the low frequency channel with their own temperatures, heat capacity and thermal conductivity. In this model, the cross-plane thermal conductivity at the equilibrium limit ($k_z^{eq}$) can be calculated by the summation of the thermal conductivity of both channels, indicated by the dash lines in Figure 2b. (See Supporting Information S3 to S5 for details of fitting, uncertainty analysis and division of phonon channels). After $k_z^{eq}$ is determined, the laser spot is tightly focused to $w = 4.2$ μm, so that the TDTR signal becomes sensitive to both the cross-plane thermal conductivity $k_z^{eq}$ and the in-plane thermal conductivity $k_r$. As shown in Figure 2c, the in-plane thermal conductivity of WSe$_{1.2}$Te$_{0.8}$ is measured to $k_r = 10.1$ W/mK with a spot radius of 4.2 μm and a modulation frequency of 0.353 MHz.

Using a similar method as described above, we have measured the in-plane and cross-plane thermal conductivity of all TMD samples as summarized in Figure 3a and Figure 3b, respectively. We compare the thermal conductivity of the pure WSe$_2$ and WTe$_2$ crystals with the results available in the literature. For WSe$_2$, our measurements for both $k_r = 40$ W/mK and $k_z = 2.45$ W/mK are consistent with our previous work[39] and the first principles calculations by Lindroth et al. [43] However, the measured $k_z$ is significantly higher than the TDTR measurements by Chiritescu *et al.* [44] and Murato *et al.*, [45] probably because they neglected non-equilibrium transport. For WTe$_2$, our measurement for the in-plane thermal conductivity ($k_r = 13.5$ W/mK) agrees well with the measurement by Zhou *et al.* ($k_r = 15$ W/mK). [46] Interestingly, the $k_r$ of WTe$_2$ measured by TDTR is even higher than the first principles calculation of lattice thermal conductivity. There are two possible reasons for such discrepancy. First of all, the electrons might have a notable contribution to the total thermal conductivity measured by the TDTR since WTe$_2$ is a semi-metal,[47] while it was completely neglected by the first principle calculations of lattice thermal conductivity.

The other possible reason is likely due to the computational error. WTe$_2$ has a rather complicated atomic structure. However, in the calculation by Liu *et al.*,[48] the authors used a relative small supercell (2×2×1 unit cells) for calculating the force constants to mitigate the large computation cost, which might lead to an underestimated thermal conductivity.[43] In the cross-plane direction, our measurement is $k_z = 1.34$ W/mK, which agrees well with both the first principles calculations[48] and the measurements by others.[46, 49-50] For both the in-plane and the cross-plane direction, the thermal conductivity reduces as the composition fraction $x$ or $(1-x)$ in WSe$_{2(1-x)}$Te$_{2x}$ approaches 0.5. The sharp change of thermal conductivity in both in-plane and cross-plane directions are observed as $x$ increases from 0.4 to 0.6. This is due to the phase transition from the 2H to Td phase as $x$ increases from 0.4 to 0.6. Since phase transition has also been used for enhancing thermoelectric figure of merit (ZT) for other calcogenides like SnSe[51], Cu$_2$Se[52] and so on, such a phase transition of 2H to Td could potentially result in a higher ZT factor due to the decreased thermal conductivity both along the in-plane direction and the cross-plane direction.

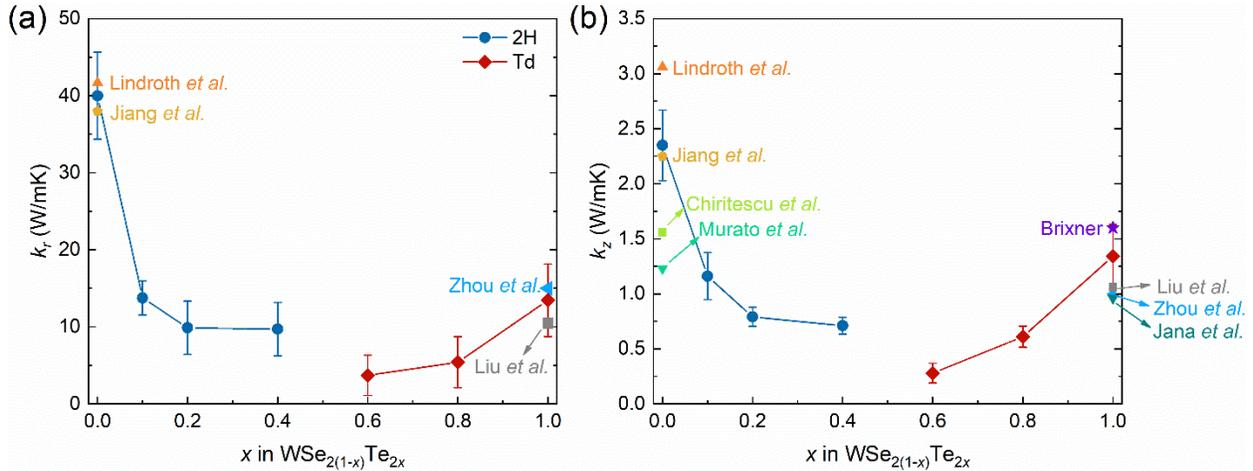

Figure 3. Composition dependent (a) In-plane thermal conductivity and (b) cross-plane thermal conductivity of WSe$_{2(1-x)}$Te$_{2x}$ measured by TDTR. The thermal conductivity of pristine WSe$_2$ ($x=0$) and WTe$_2$ ($x=1$) are compared with first-principles calculations by Lindroth *et al.*[43] and Liu *et al.*,[48] and measurement by Jiang *et al.*,[39] Chiritescu *et al.*,[44] Murato *et al.*[45], Zhou *et al.*,[46] Brixner *et al.*[49] and Jana *et al.*[50]

In Figure 4, we show the temperature-dependent $k_r$ and $k_z$ of the 2H WSe$_{1.2}$Te$_{0.8}$ (*x*=0.4) and Td WSe$_{0.4}$Te$_{1.6}$ (*x*=0.8), in comparison with the pure 2H WSe$_2$ and Td WTe$_2$, respectively. In the entire temperature range from 80 K to 300 K, both $k_r$ and $k_z$ of 2H WSe$_{1.2}$Te$_{0.8}$ are greatly reduced compared with the pure WSe$_2$ due to the alloy scattering, and a similar reduction is also observed in WSe$_{0.4}$Te$_{1.6}$ in the Td phase compared with WTe$_2$. As shown in Figure 4a and Figure 4c, the $k_r(T)$ curve flattens below 150 K for both 2H WSe$_{1.2}$Te$_{0.8}$ and Td WSe$_{0.4}$Te$_{1.6}$, while $k_r$ kept increasing with the temperature for the pristine WSe$_2$ and WTe$_2$. Since the alloy scattering due to mass disorder is temperature independent,[53] it becomes more dominant at lower temperature when the intrinsic three phonon scattering is much weaker at cryogenic temperature, contributing to the weaker temperature dependence of $k_r$ below 150 K. It is interesting to note that the $k_z(T)$ curve of WSe$_{0.4}$Te$_{1.6}$ is much flatter than the $k_z(T)$ curve WTe$_2$, as shown Figure 4d. Since WSe$_{0.4}$Te$_{1.6}$ has randomly distributed Se defects, the distribution pattern of the defects is expected to be very different in each monolayer, which breaks the periodicity in the cross-plane direction. As a result of such atomic disorder, the vibration modes are greatly localized in the cross-plane direction, resulting in the much lower $k_z$ value and a much flatter $k_z(T)$ curve than the prinstine TMD crystal. For the same reason, the $k_z(T)$ curve of WSe$_{1.2}$Te$_{0.8}$ showed a similar shape compared to WSe$_{0.4}$Te$_{1.6}$, as shown in Figure 4b and Figure 4d. However, the temperature dependence of $k_z$ of WSe$_2$ is very different from WTe$_2$. Instead of increasing with decreasing temperature, the $k_z(T)$ curve of WSe$_2$ showed a peak near 150 K. Suggested by the Lindroth et al.[43] and Jiang et al.,[39] there could exist staking faults that induce boundary scattering on the length scale of ~150 nm,[43] resulting in such a peak in the $k_z(T)$ curve and much lower $k_z$ compared with first principles calculations. However, such peak in the $k_z(T)$ curve is absent in the WTe$_2$. Such different behavior of temperature dependent $k_z$ can be explained by the different distribution of phonon mean free paths (MFPs) in WTe$_2$ and WSe$_2$. As suggested by the first principles calculations, half of $k_z$ of WTe$_2$ is contributed by phonons with MFPs between 200 nm to 1 μm,[48] but 50% of $k_z$ of WSe$_2$ is contributed by phonons with MFPs above 1 μm.[43] Due to the much longer MFPs in WSe$_2$ than WTe$_2$, the stacking faults are expected to have a much more pronounced effect on $k_z$ in WSe$_2$ than WTe$_2$.

In summary, we applied the ultrafast laser based TDTR technique to study the temperature-dependent anisotropic thermal transport of layered 2D TMD alloys WSe$_{2(1-x)}$Te$_{2x}$. We observed that the cross-plane thermal conductivity depends on the modulation frequency of the TDTR pump beam, due to the non-equilibrium transport between different phonon modes in the cross-plane direction. A two-channel heat conduction model is used to extract the cross-plane thermal conductivity at the thermal equilibrium limit. The in-plane thermal conductivity is then determined using a tightly focused laser spot with radius of 4.2 μm. Both the in-plane and cross-plane thermal conductivity is reduced at higher alloy mixing level as the composition fraction $x$ or (1-$x$) in WSe$_{2(1-x)}$Te$_{2x}$ approaches 0.5. A clear discontinuity in both the cross-plane and the in-plane thermal conductivity is observed as $x$ increases from 0.4 to 0.6 due to the phase transition from the 2H to Td phase in the layered 2D alloys. We also found that the temperature dependence of thermal conductivity for the TMD alloys becomes weaker compared with the pristine 2H WSe$_2$ and Td WTe$_2$ due to the atomic disorder. This work serves as an important starting point for exploring phonon transport physics in layered 2D alloys.

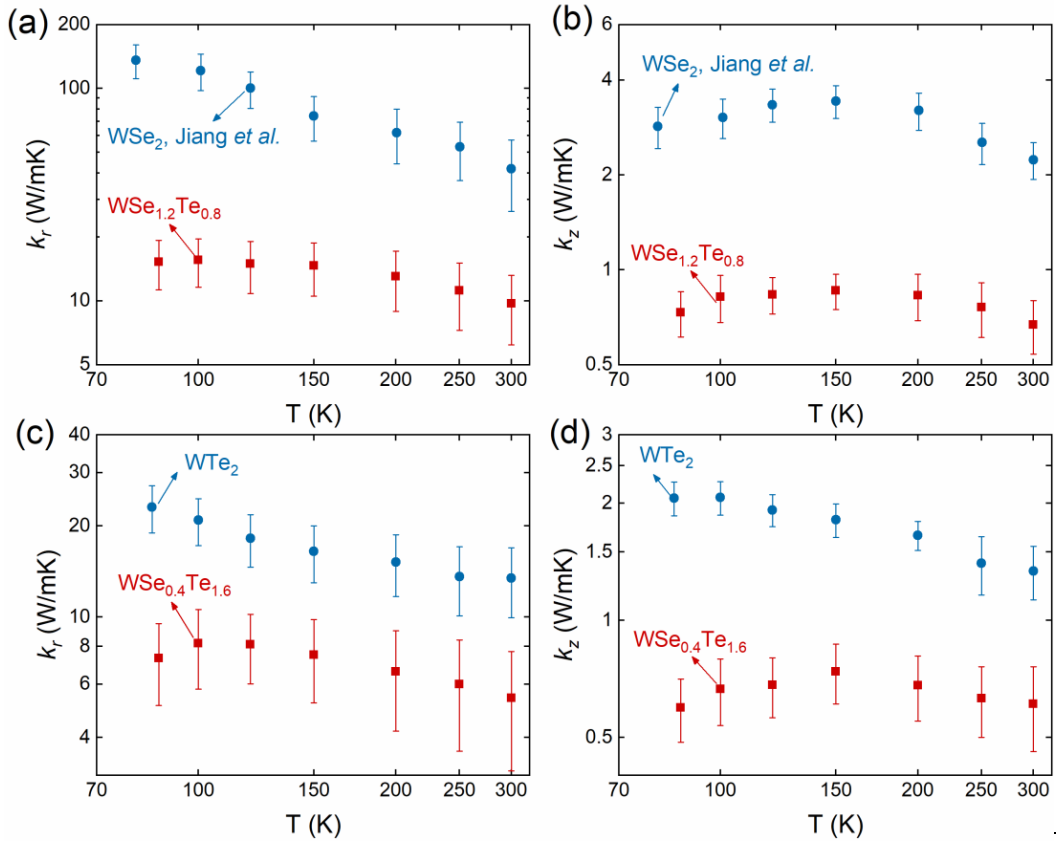

Figure 4. Temperature dependent (a) in-plane thermal conductivity and (b) cross-plane thermal conductivity of 2H $WSe_2$ and $WSe_{1.2}Te_{0.8}$ ($x$= 0.4), and temperature dependent (c) in-plane thermal conductivity and (d) cross-plane thermal conductivity of $WTe_2$ and $WSe_{0.4}Te_{1.6}$ ($x$= 0.8).

**Acknowledgment**: This work is supported by NSF (Grant No. 1512776). X. Q. acknowledges the helpful discussions with Jian Zhou and Gang Liu at Nanjing University. This work is also supported by the National Research Foundation Singapore under NRF Award No. NRF-RF2013-08.

# Supporting Information

**Anisotropic Thermal Transport in Phase-Transition Layered 2D Alloys WSe$_{2(1-x)}$Te$_{2x}$**

Xin Qian[1#], Puqing Jiang[1#], Peng Yu[2#], Xiaokun Gu[3], Zheng Liu[2*] and Rongguo Yang[1*]

[1]Department of Mechanical Engineering, University of Colorado, Boulder, CO 80309

[2]Center for Programmable Materials, School of Materials Science & Engineering, Nanyang Technological University, 649798, Singapore

[3]Institute of Engineering Thermophysics, Shanghai Jiao Tong University, Shanghai, 200240

[*]Email: Ronggui.Yang@Colorado.Edu

**Table of Contents**

**S1. Preparation of TMD alloy samples**

**S2. Non-equilibrium phonon transport and the two-channel heat conduction model**

**S3. Data reduction of TDTR measurement using the two-channel model**

**S4. Uncertainty analysis**

**S5. Heat capacity of TMD alloys using the first principles calculations**

# Supporting Information

## S1. Preparation of TMD alloy samples

Large, well-formed, single crystals of WSe$_{2(1-x)}$Te$_{2x}$ (x = 0–1) alloys were grown by chemical vapor transport (CVT) method with iodine (I) as the transporting gas. Stoichiometric amounts of tungsten (W) powder (99.9%, Sigma-Aldrich), selenium (Se) powder (99.95%, Sigma-Aldrich) and tellurium (Te) powder (99.95%, Sigma-Aldrich) with a total weight of 300 mg, plus an extra 40 mg of I as the transporting gas were sealed in an evacuated 20 cm long quartz tube under the vacuum at 10$^{-6}$ Torr. The quartz tube was placed in a three-zone furnace. Firstly, the reaction zone was maintained at 850 °C for 30 h with the growth zone at 900 °C in order to prevent the transport of the product and a complete reaction; then the reaction zone was heated to 1010 °C and held for 7 days with the growth zone at 900 °C. Finally, the furnace was naturally cooled down to the room temperature and the single crystals were collected in the growth zone. Residuals were cleaned using acetone before measurement.

After preparing the TMD alloys, we collected the powder XRD patterns using a Rigaku DMAX 2500 diffractometer with monochromatized Cu-K$\alpha$ radiation at room temperature in the 2$\theta$ range of 5–50° with a scan step width of 0.05°. SEM and semi-quantitative microprobe analyses on WSe$_{2(1-x)}$Te$_{2x}$ (x = 0–1) alloys were performed with the aid of a field emission scanning electron microscope (FESEM, JSM-5410) equipped with an energy dispersive X-ray spectroscope (EDX, Oxford INCA).

## S2. Non-equilibrium phonon transport and the two-channel heat conduction model

The frequency dependent $k_z$ can be attributed to the non-equilibrium transport between the high frequency and low frequency phonons. In layered materials like TMDs, black phosphorus and TMD alloys, low frequency phonons contribute to the majority of the $k_z$, but only a small amount of the heat capacity. In contrast, the high frequency phonons have small contribution to the $k_z$ due to their small group velocity, but contribute to the majority of the heat capacity of the material. In TMD materials, the thermal conductivity contributed by high frequency phonons is further suppressed because of the alloy scattering. Such a mismatch of the heat capacity and thermal conductivity between different phonon groups would result in a temperature difference when the sample is periodically heated by the laser. The heating would accumulate within the high frequency phonons because they dissipate much slower than the low frequency phonons with long mean free paths. To capture the non-equilibrium transport between different phonon groups within the TMD alloys, we use the two-channel heat conduction model as previously used to explain the thermal transport in TMD crystals and black phosphorus.[1-2] This model assumes that the heat carriers can be divided into two groups, and each group satisfies the Fourier's heat conduction equation while $g$ is the coupling strength between the two channels:

$$C_1 \frac{\partial T_1}{\partial t} = \frac{k_{r1}}{r} \frac{\partial}{\partial r}\left(r \frac{\partial T_1}{\partial r}\right) + k_{z1} \frac{\partial^2 T_1}{\partial z^2} + g(T_2 - T_1)$$
$$C_2 \frac{\partial T_2}{\partial t} = \frac{k_{r2}}{r} \frac{\partial}{\partial r}\left(r \frac{\partial T_2}{\partial r}\right) + k_{z2} \frac{\partial^2 T_2}{\partial z^2} + g(T_1 - T_2)$$
(1)

# Supporting Information

Here $T_i, C_i, k_{ri}, k_{zi}$ are the temperature, the heat capacity, the in-plane thermal conductivity and the cross-plane thermal conductivity of $i$-th channel ($i = 1,2$), respectively. As shown in Figure S1, channel 1 in the Al transducer represents the electrons, and channel 2 in Al transducer represents the phonons. In the TMD alloys which is modeled as a substrate, channel 1 represents the low-frequency phonons and channel 2 represents high frequency phonons. Since there are two channels on each side of the Al/TMD alloys interface, the energy transport across the interface is described by a 2-by-2 matrix $[G_{ij}]$, with the elements $G_{ij}$ meaning the conductance between channel $i$ in Al and channel $j$ in the TMD alloy. A detailed solution of the two-channel model can be found in ref. [1] and [2]. The division of phonon channels for TMD alloys are shown in Section S6.

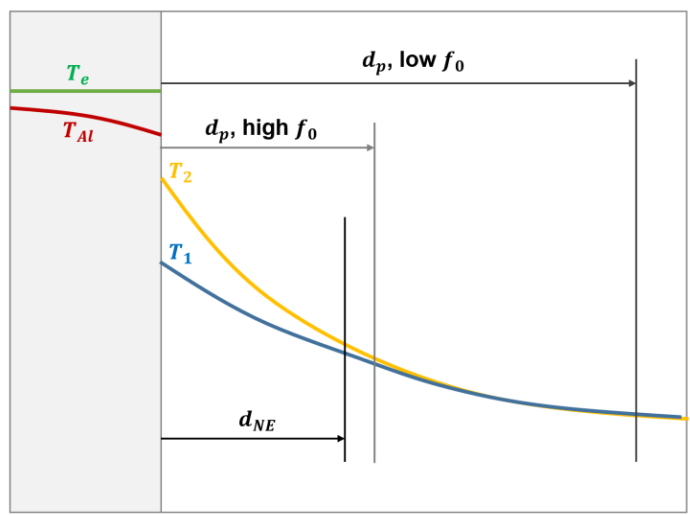

Figure S1. Schematic of the two-channel model for the non-equilibrium phonon transport.

Figure S1 illustrates that the neglect of non-equilibrium transport between high frequency and low frequency phonons in the TMD alloy substrate would result in frequency dependent $k_z$. The temperature difference between the two phonon channels is not negligible within a length scale $d_{NE} = \left(\frac{g}{k_{z1}} + \frac{g}{k_{z2}}\right)^{-1/2}$. The frequency dependence of the thermal conductivity $k_z$ is caused by the neglect of such non-equilibrium transport when the conventional one-channel heat conduction model is used for data reduction. In the one-channel model, the length scale of temperature field in the sample is characterized by the penetration depth $d_p = \sqrt{k_z/\pi C f_0}$ with $C$ the heat capacity and $f_0$ the modulation frequency, and TDTR measures the thermal property within the length scale of $d_p$. At high modulation frequency $f_0$, $d_p$ is comparable to $d_{NE}$ and the non-equilibrium transport is not negligible. As a result, $R_{NE}$ is grouped in to $k_z$ when the one-channel model is used for data reduction, which would result in an underestimated $k_z$. When the $f_0$ is low, $d_p$ is larger than $d_{NE}$, and the non-equilibrium resistance has less effect on $k_z$, thus $k_z$ increases when $f_0$ decreases. For example, the non-equilibrium length of WSe$_{1.2}$Te$_{0.8}$ is estimated as $d_{NE} \approx 90$ nm, estimated from the parameters in Table S1. At high modulation frequency $f_0 = 9.8$ MHz, the

# Supporting Information

penetration depth $d_p \approx 110$ nm, which is comparable to $d_{NE}$. When $f_0 = 0.353$ MHz, $d_p \approx 600$ nm, and the non-equilibrium transport has less effect on $k_z$.

## S3. Data reduction of TDTR measurement using the two-channel model

For the transducer layer, the electronic thermal conductivity of Al is estimated to be 150 W/mK by using a four-point probe for electrical conductivity measurement, which is used as the value of $k_{r1}$ and $k_{z1}$. The phononic thermal conductivity $k_{r2} = k_{z2} \approx 6$ W/mK, is estimated from first-principles calculations.[3] The electron phonon coupling strength in Al transducer is taken as $g = 24.5 \times 10^{16}$ W/m³K.[4] Since interfacial transport through the electron-phonon channel across the interface is usually negligible,[1] we apply adiabatic boundary condition for the electron channel in the Al transducer, so that $G_{11} = G_{12} = 0$.

The remaining unknown parameters of the TMD layers $\boldsymbol{U} = [k_{r1}, k_{r2}, k_{z1}, k_{z2}, G_{21}, G_{22}, g]^T$ need to be determined by matching the TDTR experimental signals. We conduct the sensitivity analysis so that in-plane properties $k_{r1}, k_{r2}$ and the cross-plane properties can be measured separately. The sensitivity of the signal with respect to the parameter $p$ is defined as:

$$S_p = \frac{\partial \ln\left(-\frac{V_{in}}{V_{out}}\right)}{\partial \ln p} \tag{2}$$

where $-\frac{V_{in}}{V_{out}}$ is the ratio between the in-phase signal $V_{in}$ and the out-of-phase signal $V_{out}$. Figure S2a shows the sensitivity analysis based on the two channel heat conduction model. At relatively large spot radius $w = 8.4$ µm above 1 MHz, the five parameters $[k_{z1}, k_{z2}, G_{21}, G_{22}, g]$ dominantly affect the signal $-V_{in}/V_{out}$. These five parameters are extracted simultaneously by fitting the experiments at five modulation frequencies: $f_0 = 1.06$ MHz, 2.16 MHz, 5.10 MHz and 9.80 MHz. To fit the obtained experiment signal $-V_{in}/V_{out}$, we use the nonlinear regression method[5] to minimize the cost function, which is defined as:

$$W(\boldsymbol{U}) = \sum_i \sum_j \left[R_{Exp}(\tau_i, f_{0j}) - F(\tau_i, f_{0j}, \boldsymbol{U}, \boldsymbol{P})\right]^2 \tag{3}$$

where $R_{Exp}(\tau_i, f_{0j})$ is the ratio $-V_{in}/V_{out}$ measured experimentally at the delay time $\tau_i$ and the modulation frequency $f_{0j}$. $F$ is the solution of the two-channel heat conduction model which predicts the signal $-V_{in}/V_{out}$. The vector $\boldsymbol{U} = [k_{z1}, k_{z2}, G_{21}, G_{22}, g]^T$ is the set of unknown parameters that need to be determined. The vector $\boldsymbol{P}$ is the vector of control parameters including laser spot radius, thickness, heat capacity and thermal conductivity of the transducer, and heat capacity of the sample. The Simplex algorithm[5] is used to seek the minimum of the cost function by varying the values of $\boldsymbol{U}$ iteratively, until the change in $\boldsymbol{U}$ and $W$ is both smaller than 0.1%.

At the local equilibrium limit, the cross-plane thermal conductivity is simply calculated as the summation of the contribution from each channel:

$$k_z^{eq} = k_{z1} + k_{z2} \tag{4}$$

# Supporting Information

The sensitivity for the in-plane thermal conductivity is increased from 0.05 to 0.09 when the laser spot is tightly focused to $w = 4.2$ μm at low frequency $f = 0.3$ MHz, as shown Figure S2b. Therefore, $f_0 = 0.353$ MHz is selected for measuring $k_r$ in the main text. After the cross-plane transport properties $[k_{z1}, k_{z2}, G_{21}, G_{22}, g]$ are determined, the unknown parameter is set to be $\boldsymbol{U} = [k_{r1}, k_{r2}]^\text{T}$ and the cross-plane transport properties are grouped into the vector of control variables $\boldsymbol{P}$ which are fixed during the nonlinear regression. $k_{r1}, k_{r2}$ are then extracted by minimizing Eq. (3). The in-plane thermal conductivity at the near equilibrium limit is similarly calculated by the summation of the contributions from both channels:

$$k_r^{eq} = k_{r1} + k_{r2} \tag{5}$$

Table S1 reports the best-fit parameters $[k_{r1}, k_{r2}, k_{z1}, k_{z2}, G_{21}, G_{22}, g]$ and the heat capacities of each channel of the TMD alloys. The division of heat capacity into the two channels are described in Section S5.

Table S1. Best-fit parameters $[k_{r1}, k_{r2}, k_{z1}, k_{z2}, G_{21}, G_{22}, g]$ and the heat capacity of the TMD samples WSe$_{2(1-x)}$Te$_{2x}$. Shaded area represents the samples in Td phase while the rest are the samples in the 2H phase.

| $x$ | $C_1$ MJ/m³K | $C_2$ MJ/m³K | $k_{r1}$ W/mK | $k_{r2}$ W/mK | $k_{z1}$ W/mK | $k_{z2}$ W/mK | $G_{21}$ MW/m²K | $G_{22}$ MW/m²K | $g$ 10¹³W/m³K |
|---|---|---|---|---|---|---|---|---|---|
| 0 | 0.693 | 1.287 | 32.1 | 7.2 | 2.05 | 0.4 | 7.15 | 46.3 | 8.0 |
| 0.1 | 0.640 | 1.300 | 10.62 | 3.14 | 0.94 | 0.23 | 1.38 | 34.6 | 4.4 |
| 0.2 | 0.623 | 1.277 | 9.84 | 1.03 | 0.71 | 0.08 | 4.21 | 9.81 | 1.1 |
| 0.4 | 0.613 | 1.227 | 9.72 | 0.34 | 0.65 | 0.06 | 3.99 | 8.97 | 0.65 |
| 0.6 | 0.423 | 1.377 | 3.2 | 0.5 | 0.24 | 0.04 | 26.2 | 34.2 | 0.73 |
| 0.8 | 0.439 | 1.301 | 4.8 | 0.6 | 0.51 | 0.10 | 29.2 | 38.7 | 0.8 |
| 1.0 | 0.443 | 1.237 | 10.41 | 4.64 | 1.14 | 0.21 | 23.2 | 43.1 | 3.3 |

# Supporting Information

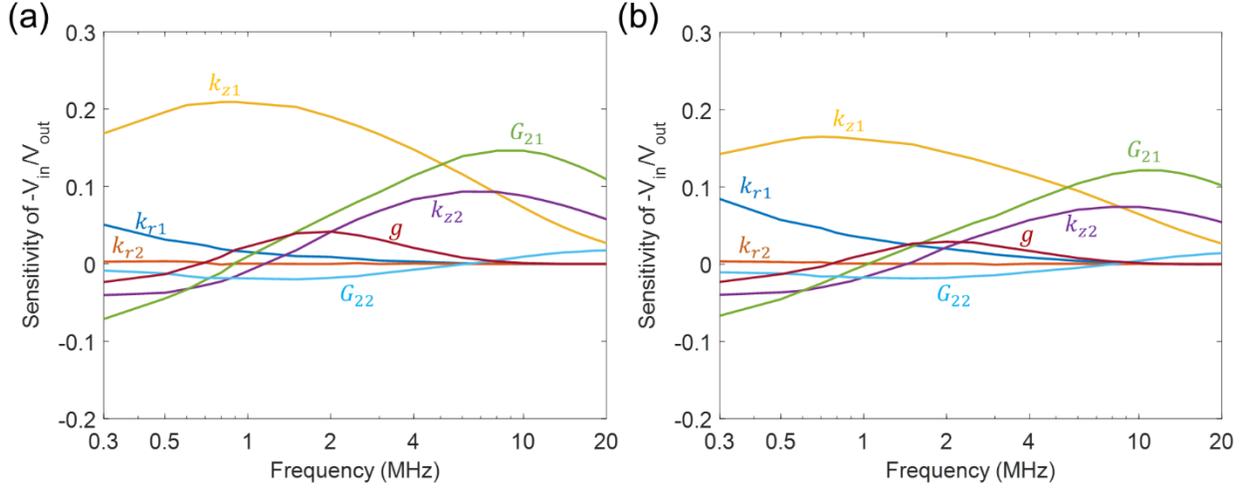

Figure S2. Sensitivity analysis of experimental parameters for a 76 nm Al on WSe$_{1.2}$Te$_{0.8}$ sample as a function of modulation frequency $f_0$. (a) Sensitivity of the unknown parameters using a $w = 8.4$ μm and (b) $w = 4.2$ μm. The parameters for calculating sensitivities are $k_{r1} = 9.7$ W/mK, $k_{r2} = 0.3$ W/mK, $k_{z1} = 0.65$ W/mK, $k_{z2} = 0.05$ W/mK, $G_{21} = 4.0$ MW/m$^2$K, $G_{22} = 8.97$ MW/m$^2$K and $g = 6.5 \times 10^{12}$ W/m$^3$K.

**S4. Uncertainty analysis.**

The uncertainty of the unknown variables $U$ is estimated using the following propagation formula:[2, 6]

$$var[U] = \Sigma_U^{-1} \left[ \sum_j J_U^T(f_{0j}) var[R_{Exp}(f_{0j})] J_U(f_{0j}) \right] \Sigma_U^{-1} + \Sigma_U^{-1} \Sigma_{UP} var[P] \Sigma_{UP}^T \Sigma_U^{-1} \quad (6)$$

where $var[\cdot]$ denotes the variance matrix, $J_U(f_{0j}) = \left(\frac{\partial F}{\partial U}\right)_{f_{0j}}$ and $J_P(f_{0j}) = \left(\frac{\partial F}{\partial P}\right)_{f_{0j}}$ are the Jacobi matrices of the vector of model prediction $F = [F(\tau_1), F(\tau_2), \ldots, F(\tau_i), \ldots]_{f_{0j}}$ with respect to the unknown parameters $U$ and the control parameters $P$. The $\Sigma$ matrices are written as:

$$\Sigma_U = \sum_j J_U^T(f_{0j}) J_U(f_{0j}), \quad \Sigma_{UP} = \sum_j J_U^T(f_{0j}) J_P(f_{0j}) \quad (7)$$

Since uncertainties of the control parameters are uncorrelated, the covariance matrix $var[P]$ is written as a diagonal matrix:

$$var[P] = diag[\sigma_{k_{Al}}^2, \sigma_{C_{Al}}^2, \sigma_{d_{Al}}^2, \sigma_C^2, \sigma_w^2] \quad (8)$$

where $\sigma$ is the standard deviation. The uncertainties of the input parameters $P$ are 10% for the thermal conductivity of Al, 3% for the heat capacity of Al and the substrate, 4% for the Al thickness, and 3% for the laser spot size.

# Supporting Information

The term $\Sigma_U^{-1}[\sum_j J_U^T(f_{0j}) var[R_{Exp}(f_{0j})] J_U(f_{0j})]\Sigma_U^{-1}$ describes the uncertainty contributed by the noise of the signal, where the $var[R_{Exp}(f_{0j})]$ represents the noise of the signal at the frequency $f_{0j}$, which is obtained by calculating the variance of the signal $-V_{in}/V_{out}$ among five individual measurements at each sampled delay time $\tau_{0j}$. The experimental noise only contributes to less than 5% of the uncertainty, and the majority of the error comes from the uncertainties of the control variables. If $U$ is a vector of $N$ elements $U = [u_1, u_2, ..., u_N]^T$, the covariance matrix $var[U]$ is a $N \times N$ matrix with the following form:

$$var[U] = \begin{bmatrix} \sigma_{u_1}^2 & cov[u_1, u_2] & \cdots & cov[u_1, u_N] \\ cov[u_2, u_1] & \sigma_{u_2}^2 & \cdots & cov[u_2, u_N] \\ \vdots & \vdots & \ddots & \vdots \\ cov[u_N, u_1] & cov[u_N, u_2] & \cdots & \sigma_{u_N}^2 \end{bmatrix} \quad (9)$$

where $cov[u_i, u_j]$ is the covariance between $u_i$ and $u_j$ and it is identical to $cov[u_j, u_i]$ so that the $var[U]$ is a symmetrical matrix. The covariance $cov[u_i, u_j]$ denotes the correlation between the two variables $u_i$ and $u_j$. If $cov[u_i, u_j]$ is zero, then $u_i$ and $u_j$ are independent. When determining the confidence interval of the multiple parameters, it is also necessary to consider the covariance. The confidence interval for multiple parameters are determined by a quadratic surface in the parameters space:

$$(U - U^0)^T (var[U])^{-1} (U - U^0) = \chi_N^2 (P = 0.95) \quad (10)$$

where $U^0$ denotes the best-fit parameters. $\chi_N^2(P)$ is the $N$-th order quantile function,[7] and $P = 0.95$ is the probability of the confidence interval. To estimate the error of $k_z^{eq}$ and $k_r^{eq}$, we first plot the confidence interval projected to the sub-space of the entire parameter space using the following two equations respectively:

$$([k_{r1}, k_{r2}] - [k_{r1}^0, k_{r2}^0])(var[k_{r1}, k_{r2}])^{-1}([k_{r1}, k_{r2}] - [k_{r1}^0, k_{r2}^0])^T = \chi_2(0.95) \quad (11)$$

and

$$([k_{z1}, k_{z2}] - [k_{z1}^0, k_{z2}^0])(var[k_{z1}, k_{z2}])^{-1}([k_{z1}, k_{z2}] - [k_{z1}^0, k_{z2}^0])^T = \chi_2(0.95). \quad (12)$$

The confidence intervals are generally ellipses as shown in Figure S3. Based on the shape of the confidence interval, the upper and lower limit of $k_z^{eq}$ and $k_r^{eq}$ can be obtained as showed in Figure S3. For example, when determining $k_z^{eq}$, the line tangential to the ellipse with the equation:

$$k_{z1} + k_{z2} = k_z^{eq} \quad (13)$$

are drawn, and we can determine the uncertainty of $k_z^{eq}$ from the intersect with the two axes.

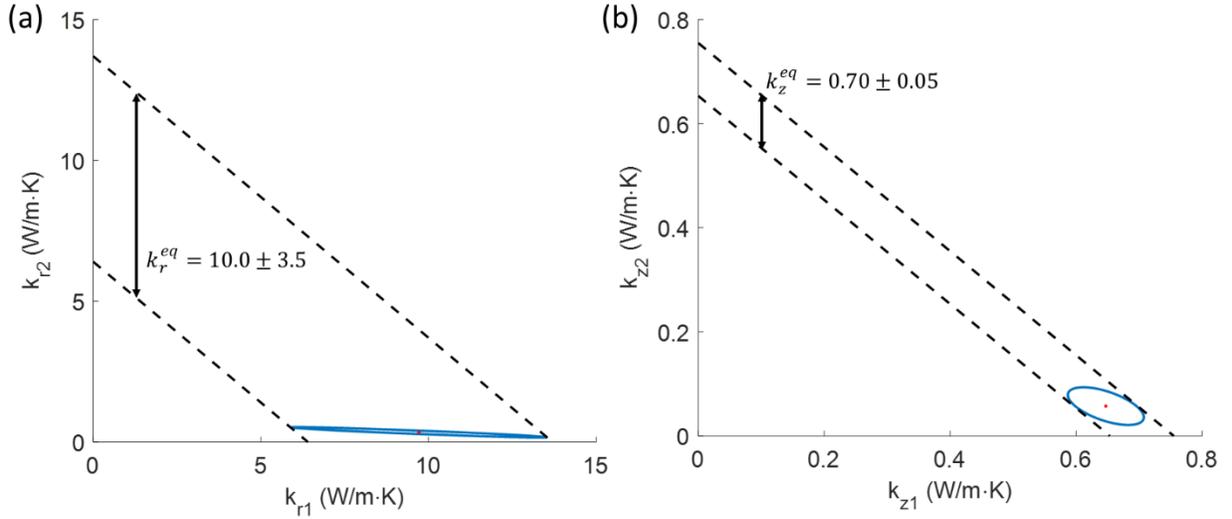

Figure S3. Schematic of the confidence intervals (ellipses with blue solid lines) and the determination of the thermal conductivity at the near equilibrium limit from the confidence interval (dashed black lines). The red dot shows the best fit value. (a) confidence interval of $k_{r1}$ and $k_{r2}$. (b) confidence interval for $k_{z1}$ and $k_{z2}$ for WSe$_{1.2}$Te$_{0.8}$.

**S5. Heat capacity of TMD alloys using the first principles calculations**

*Virtual crystal approximation for calculating heat capacities.* We use the first principles lattice dynamics calculations to determine the heat capacities of TMD alloys. For crystals, the heat capacity could be easily calculated from the phonon dispersions. However, phonon dispersion is not well defined due to the lack of periodicity in the layered TMD alloys with randomly distributed Se and Te atoms on the chalcogen sites. We therefore use the virtual crystal approximation (VCA) to restore the periodicity so that phonon dispersions of alloys can be calculated.[8] Figure S4 describes the application of VCA to the TMD alloys in 2H phase as an example. A virtual crystal is created by replacing the randomly distributed Se and Te atoms with a virtual atom X, whose atomic mass ($m_X$), lattice constants ($a$ and $c$) and harmonic force constants $\boldsymbol{\phi}$ are a compositional average of 2H WSe$_2$ and 2H WTe$_2$:

$$\begin{aligned}
m_X &= xm_{\text{Te}} + (1-x)m_{\text{Se}} \\
a &= xa_{\text{WSe}_2} + (1-x)a_{\text{WSe}_2} \\
c &= xc_{\text{WSe}_2} + (1-x)c_{\text{WSe}_2} \\
\boldsymbol{\phi} &= x\boldsymbol{\phi}_{\text{WTe}_2} + (1-x)\boldsymbol{\phi}_{\text{WSe}_2}
\end{aligned} \quad (14)$$

where $x$ is the fraction of WTe$_2$ in the alloy WSe$_{2(1-x)}$Te$_{2x}$. Similarly, the VCA is applied to obtain the lattice constants and harmonic force constants of TMD alloys in the Td phase.

# Supporting Information

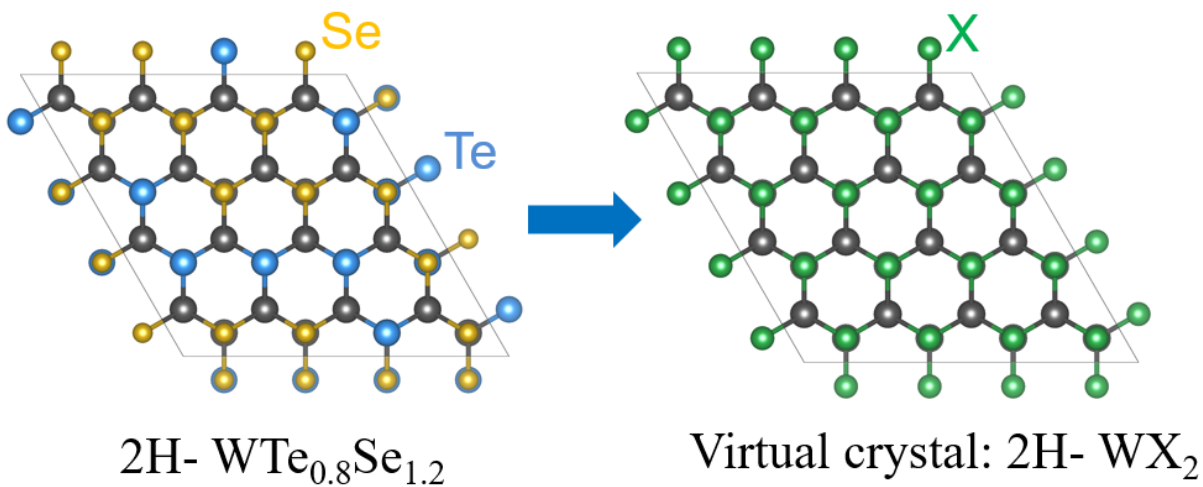

Figure S4. Schematic of virtual crystal approximation for TMD alloys in the 2H phase.

*First-principles calculations of the harmonic force constants*. First-principles calculations are carried out using Vienna ab-initio Simulation Package (VASP)[9-10] with the projector augmented wave (PAW) pseudopotential[10] with the local density approximation (LDA). The energy cut-off for the plane-wave basis is set to be 600 eV for all calculations. A $13 \times 7 \times 3$ Monkhorst-Pack mesh (k-mesh) is used to sample the Brillouin zone during the structure optimization. The choice of the energy cutoff and k-mesh ensures that the energy change is smaller than 1 meV/atom when refining these two parameters.[11] All materials are relaxed through the conjugate gradient algorithm until the atomic forces are smaller than $5\times10^{-6}$ eV/Å. The obtained lattice parameters after the structural relaxation are shown in Table S2, all agreeing well with the literature values[12-13] within 2%.

Table S2. Lattice parameters of $WSe_2$ and $WTe_2$ compared with literature values.

|   | 2H $WSe_2$ | | 2H $WTe_2$ | |
|---|---|---|---|---|
|   | This work | Exp.[a] | This work | Calc. (LDA)[b] |
| $a$ | 3.247 | 3.282 | 3.472 | 3.51 |
| $c$ | 12.792 | 12.961 | 13.804 | 13.90 |
|   | Td $WSe_2$ | | Td $WTe_2$ | |
|   | This work | - | This work | Exp. |
| $a$ | 3.248 | - | 3.443 | 3.486 |
| $b$ | 5.852 | - | 6.215 | 6.265 |
| $c$ | 12.679 | - | 13.802 | 14.038 |

a. Ref. [12]; b. LDA calculation by Ref. [13]

We use the Phonopy package[14] for extracting the harmonic force constants of TMD alloys. The standard direct method is employed to extract the harmonic and third-order anharmonic force constants.[15] First, we construct a series of supercells containing $3\times3\times1$ unit cells[16] with different

# Supporting Information

atoms perturbed from the equilibrium position by a small displacement $\Delta = 0.03$ Å. Then self-consistent field calculations are performed on each perturbed supercell using VASP, with a 4×4×3 k-mesh.[16] Interatomic forces are recorded for calculating the force constants using the central differentiation method:

$$\phi_{lb,l'b'}^{\alpha\beta} = -\frac{F_{l'b'}^{\alpha}(u_{lb}^{\alpha} = \Delta) - F_{l'b'}^{\beta}(u_{lb}^{\alpha} = -\Delta)}{2\Delta} \qquad (15)$$

where $\phi$ denotes the harmonic force constants; $u$ denotes the displacement from equilibrium position; $\alpha, \beta, \gamma$ are the indices for the directions in Cartesian coordinates; $l$ is the index for a unit cell in the supercell and $b$ is the $b$-th atom in the unit cell. The phonon dispersions

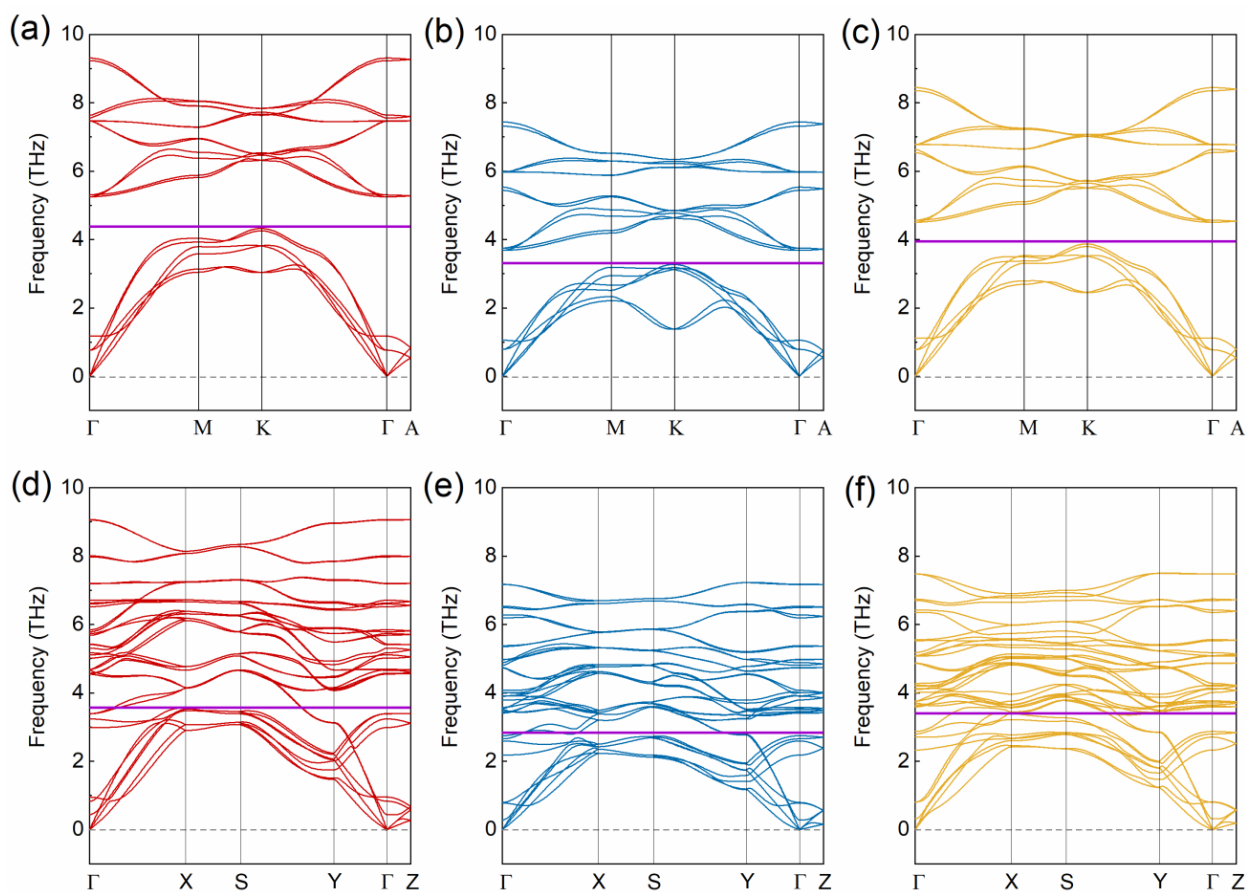

Figure S5. Phonon dispersions of (a) 2H WSe$_2$, (b) 2H WTe$_2$ and (c) 2H WSe$_{1.2}$Te$_{0.8}$, (d) Td WSe$_2$, (e) Td WTe$_2$ and (f) Td WSe$_{0.4}$Te$_{1.6}$. The horizontal purple lines show the cutoff frequencies that divide phonons into the low frequency channel and high frequency channels.

***Phonon dispersion and heat capacities of TMD alloys***. The phonon dispersion is calculated by solving the dynamical equation:

# Supporting Information

$$\sum_{b'\beta} D_{bb'}^{\alpha\beta}(\boldsymbol{q}) e_{b'}^{\beta}(\boldsymbol{q}s) = [\omega(\boldsymbol{q}s)]^2 e_b^{\alpha}(\boldsymbol{q}s) \quad (16)$$

where $b$ and $b'$ are the indices of basis atoms in the unit cell, $\alpha$ and $\beta$ are the indices of Cartesian coordinates, $\boldsymbol{q}$ is the phonon wavevector and $s$ is the polarization, $\omega$ and $e$ denote the frequency and wave vector, respectively. The dynamic matrix $D_{bb'}^{\alpha\beta}(\boldsymbol{q})$ is calculated using the harmonic force constants:

$$D_{bb'}^{\alpha\beta}(\boldsymbol{q}) = \frac{1}{\sqrt{m_b m_{b'}}} \sum_{l'} \phi_{0b,l'b'}^{\alpha\beta} \cdot \exp[i\boldsymbol{q} \cdot (\boldsymbol{R}(\boldsymbol{l'}))] \quad (17)$$

where $m_b$ is the mass of the $b$-th atom in the unit cell and $\boldsymbol{R}(\boldsymbol{l'})$ is the lattice vector of $\boldsymbol{l'}$-th unit cell. The calculated phonon dispersions of WSe$_2$, WTe$_2$ and TMD alloys in 2H phase and Td phase are shown in Figure S5. After solving the phonon dispersion, the heat capacities are calculated as:

$$C_V = \frac{\partial}{\partial T} \sum_{qs} \hbar\omega(\boldsymbol{q}s) \left[\frac{1}{2} + \frac{1}{\exp(\hbar\omega(\boldsymbol{q}s)/k_B T) - 1}\right] \quad (18)$$

where $T$ is the temperature, $k_B$ is Boltzmann constant and $\hbar$ is the reduced Planck constant. The calculated composition dependent heat capacity $C$ is shown Figure S6. The pure WSe$_2$ and WTe$_2$ agrees well with literature values.

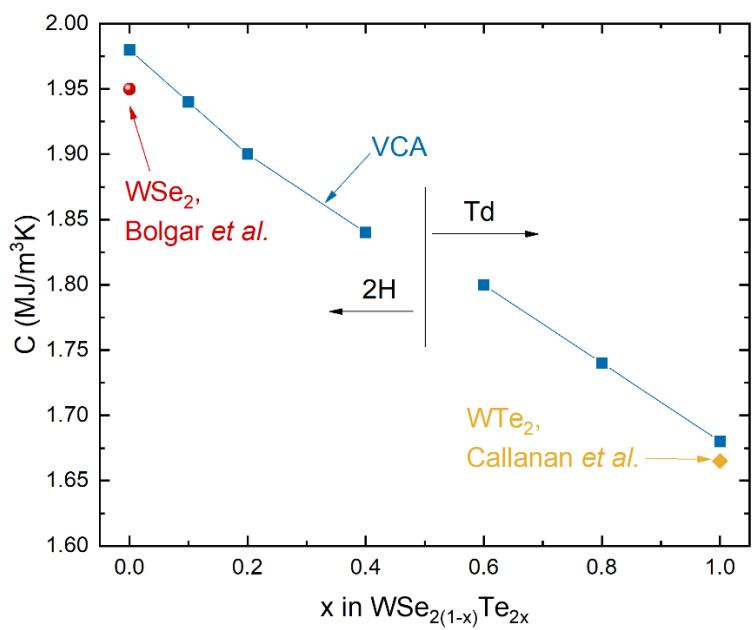

Figure S6. Composition dependent heat capacity of TMD Alloys. The heat capacity of WSe$_2$ measured Bolgar et al.[17] and the heat capacity of WTe$_2$ measured by Callanan et al.[18] are included as references.

## S6. Division of the High Frequency and the Low Frequency Channels

# Supporting Information

We divide the phonons in the TMD alloys according to the phonon dispersions obtained from VCA. For the alloys 2H phase, the heat capacity is divided into two channels according to the bandgap in the phonon dispersion:

$$C_1 = \frac{\partial}{\partial T} \sum_{\omega_{qs} < \omega_c} \hbar \omega_{qs} n_{qs}(T)$$
$$C_2 = \frac{\partial}{\partial T} \sum_{\omega_{qs} > \omega_c} \hbar \omega_{qs} n_{qs}(T)$$
(19)

where $\omega_c$ is the cutoff frequency. Based on the physical picture discussed in Section S2, the cutoff frequency $\omega_c$ should be set to include all acoustic phonons with large group velocities. Therefore, we set $\omega_c$ to the maximum frequency of LA/LO branches at Brillouin zone boundaries, as shown in Figure S5. After setting the cutoff frequency, the heat capacities can be divided into the two channels, as shown in Figure S7a-b.

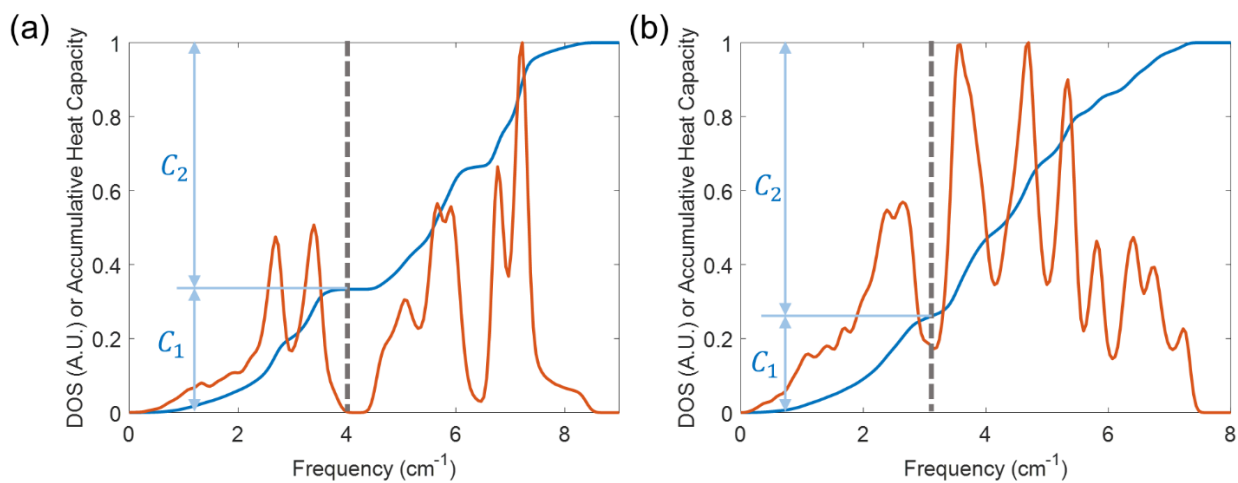

Figure S7. (a) Division of the low-frequency and the high frequency channels according to the phonon bandgap in 2H WSe$_{1.2}$Te$_{0.8}$. (b) Division of the low-frequency and the high frequency channels for Td WSe$_{0.4}$Te$_{1.6}$.

# Supporting Information